\documentstyle[12pt]{article}
\title{\bf String Representation for the\\ 
't Hooft Loop Average in the\\ 
Abelian Higgs Model}
\author{D.V.ANTONOV \thanks{E-mail addresses: 
antonov@pha2.physik.hu-berlin.de and antonov@vxitep.itep.ru}{\,}
\thanks{Supported by Graduiertenkolleg {\it Elementarteilchenphysik}, 
Russian Fundamental Research Foundation, Grant No.96-02-19184, DFG-RFFI, 
Grant 436 RUS 113/309/0, and by the INTAS, Grant No.94-2851.}
\\
{\it Institute of Theoretical and Experimental Physics,}\\
{\it B.Cheremushkinskaya 25, 117 218, Moscow, Russia}\\
{\it and}\\
{\it Institut f\"ur Physik, Humboldt-Universit\"at zu Berlin,}\\
{\it Invalidenstrasse 110, D-10115, Berlin, Germany}} 
\date{}
\begin{document}
\maketitle
\vspace{1mm}
\centerline{\bf {Abstract}}
\vspace{3mm}
Making use of the duality transformation, we derive in the 
Londons' limit of the Abelian Higgs Model string 
representation for the 't~Hooft loop average defined on the string 
world-sheet, which yields the 
values of two coefficient functions parametrizing the bilocal correlator 
of the dual field strength tensors. The asymptotic behaviours of these 
functions agree with the ones obtained within the Method of Vacuum 
Correlators in QCD in the lowest order of perturbation theory. 
We demonstrate  
that the bilocal approximation to the Method of Vacuum Correlators 
is an exact result in the Londons' limit, i.e. all the higher cumulants 
in this limit vanish. We also show that at large distances, apart 
from the integration over 
metrics, the obtained string effective theory (which in this case 
reduces to the  
nonlinear massive axionic sigma model) coincides  
with the low-energy limit of the dual version of 
4D compact QED, the so-called Universal Confining String Theory. 
We derive string tension of the Nambu-Goto term 
and the coupling constant of the rigidity term for the obtained string 
effective theory and demonstrate that the latter one is always negative, 
which means the 
stability of strings, while the positiveness of the former is confirmed 
by the present lattice data. These data enable us to find the Higgs boson 
charge and the vacuum expectation value of the Higgs field, which model 
QCD best of all. We also study dynamics of the weight factor of the 
obtained string representation for the 't~Hooft 
average in the loop space.  
In conclusion, we obtain string representation 
for the partition function of the correlators of an arbitrary number 
of Higgs currents, by virtue of which we rederive the structure  
of the bilocal correlator of the dual field strength tensors, which 
yields the surface term in the string effective action.

\newpage
{\large \bf 1. Introduction}
\vspace{3mm}   

Abelian Higgs Model (AHM) in the Londons' limit is known to display a lot 
of properties of the confining phase of gluodynamics, 
which makes it reasonable 
to consider Abrikosov-Nielsen-Olesen strings in this model as the objects 
which could tell us something about the behaviour of the 
gluodynamics string. 
To this end, in Ref. 1 the AHM partition function has been rewritten 
via string variables by making use of the duality transformation$^{2}$ and 
accounting for the Jacobian, which emerges when one passes from the 
collective field variables to the string ones. This Jacobian has been 
shown to yield the Polchinski-Strominger term$^{3}$ in the string action 
with the coupling constant ensuring exact cancellation of the conformal 
anomaly in 4D. 

An alternative approach to the string representation of gauge theories 
based on the Method of Vacuum Correlators (MVC)$^{4,5}$ has been proposed 
in Ref. 6, where the QCD string 
effective action has been derived by performing the expansion of the 
Wilson loop average. Both in Refs. 1 and 6, the sign of the rigidity 
term has been obtained to be negative, which means the stability of 
strings in the models under consideration$^{7,8}$.
However, the two main disadvantages of the method proposed in Ref. 6 
w.r.t. 
the one of Ref. 1 were the lack of the mechanism of cancellation of 
the conformal anomaly and the integration over 
metrics in the final expression for the Wilson average. The most natural 
way in the solution of these problems might lie in considering the 
strong background fields in QCD, which ensure confinement, as a 
collective variable leading to the summation over surfaces. 

In this Letter, we shall combine the approaches of Refs. 1 and 6 by 
considering the 't~Hooft loop average in the AHM without monopoles 
and find the 
string representation for it. By making use of this representation, 
we shall demonstrate that the bilocal approximation to the MVC is an 
exact result in the Londons' limit, and obtain the values of two 
coefficient functions which 
parametrize the bilocal correlator of the dual  
field strength tensors. Studying the asymptotic behaviours 
of these functions at small and large distances and comparing them 
with the corresponding behaviours known from the bilocal approximation 
to MVC in gluodynamics, 
we shall conclude that the Londons' limit reproduces  
well the long distance regime of gluodynamics, 
whereas the short distance one 
is reproduced only in the lowest order of perturbation theory. These 
points will be the topic of the next Section.

In Section 3, we shall present the values of the string tension of 
the Nambu-Goto term and the rigidity term coupling constant in the 
obtained string effective theory. The latter one occurs to be always 
negative, which, as it was mentioned above, means the stability of 
strings. Demanding the positiveness of the string tension, we arrive 
at some constraint to the Higgs boson charge, which is satisfied 
according to the present lattice data. Based on these data, we also find 
the value of the v.e.v. of the Higgs field, which corresponds to 
gluodynamics best of all.

In Section 4, we derive the equation of motion for the weight factor 
in the string representation of the 't~Hooft loop average in 
the loop space. Both this equation and the string representation 
obtained in Section 2 demonstrate that, apart from the 
integration over metrics, 
the dual version of AHM in the Londons' limit coincides at large 
distances with the large distance regime of the dual version of the 
4D compact QED, the so-called Universal Confining 
String Theory (UCST)$^{9,10}$, whose loop space dynamics was 
investigated in Ref. 11.

In Section 5, we obtain string representation for the partition 
function of the correlators of 
an arbitrary number of the Higgs currents, with the help of which we 
then get the correlator of two such currents. Due to the semiclassical 
connection 
between this correlator and the structure in the bilocal correlator 
of the dual field strength tensors which yields surface term 
in the string effective action$^{5}$, we 
rederive the coefficient function standing at this structure.

In Conclusion, we shall focus ourselves at some physical consequences 
of the obtained results and discuss possible further developments 
of the suggested approach.

\vspace{6mm}
{\large\bf 2. String Representation for the 't Hooft Loop Average 
and the Exact Result for the Bilocal Correlator of the Dual Field 
Strength Tensors in the Londons' Limit}
\vspace{3mm}

In this Section, we shall consider string representation for the 
't Hooft loop average, $\int {\cal D}x_\mu(\xi)\left<{\cal F}_N(S)
\right>$, which is defined on the closed string world-sheet $S$ (we consider 
AHM without monopoles, where the only surfaces are the 
Abrikosov-Nielsen-Olesen strings' world-sheets). 
In what follows, we shall be interested in the weight factor of 
this representation,

$$\left<{\cal F}_N(S)\right>=\int \left|\Phi\right| {\cal D}\left|
\Phi\right| {\cal D}A_\mu {\cal D}\theta^{{\rm reg.}}
\exp\Biggl\{-\int dx\Biggl[\frac14 F_{\mu\nu}^2+\frac12\left|D_\mu\Phi
\right|^2+\lambda\left(\left|\Phi\right|^2-\eta^2\right)^2+$$

$$+\frac{\pi}
{Ne}\varepsilon_{\mu\nu\alpha\beta}T_{\mu\nu}F_{\alpha\beta}\Biggr]
\Biggr\}, \eqno (1)$$
where $\Phi(x)=\left|\Phi(x)\right| {\rm e}^{i\theta(x)}$, $\theta=
\theta^{{\rm sing.}}+\theta^{{\rm reg.}}$, 
$T_{\mu\nu}(x)\equiv\int\limits_S^{} 
d\sigma_{\mu\nu}(x(\xi))\delta (x-x(\xi))$ 
is the vorticity tensor current, which is a functional of $S$, 
and $D_\mu\equiv\partial_\mu-iNeA_\mu$, 
so that the Higgs boson 
carries electric charge $Ne$. Notice, that, as it was already mentioned 
in the Introduction, in  what follows, we shall, similarly to Ref. 8, 
for simplicity disregard 
the Jacobian arising when one passes from the integration over $\theta^{{\rm 
sing.}}$ to the integration over $x_\mu(\xi)$. It is also 
worth mentioning, that from now on 
we shall make use of notations of Ref. 1, so that due to the lack of 
imaginary unit in front of the last term on the R.H.S. of Eq. (1), the 
sign of the 
second term of the cumulant expansion of Eq. (1) will be opposite to the 
standard one.    

In the Londons' 
limit, Eq. (1) takes the form

$$\left<{\cal F}_N(S)\right>=\int {\cal D}A_\mu {\cal D}
\theta^{{\rm reg.}}\exp 
\left\{-\int dx\left[\frac14 F_{\mu\nu}^2+\frac{\eta^2}{2}\left(
\partial_\mu\theta-NeA_\mu\right)^2+\frac{\pi}{Ne}\varepsilon_{\mu\nu
\alpha\beta}T_{\mu\nu}F_{\alpha\beta}\right]\right\}. \eqno (2)$$
Notice, that the $S$-dependence is present in Eq. (2) both in the 
last term and in the term $\partial_\mu\theta^{{\rm sing.}}$ due to 
the equation$^{2}$ $\varepsilon_{\mu\nu\rho\sigma}\partial_\rho 
\partial_\sigma\theta^{{\rm sing.}}(x)=2\pi T_{\mu\nu}(x)$. That is why, the 
naive application of the Gauss theorem (where only the $S$-dependence 
in the last term in Eq. (2) is accounted for) is not valid, and 
therefore the $S$-dependence in Eq. (2) could not be reduced to the 
volume one.

Performing the duality transformation of the R.H.S. of Eq. (2) along 
the lines described in Refs. 1,2, and 8, we arrive at the 
following formula

$$\left<{\cal F}_N(S)\right>=\int {\cal D}A_\mu 
{\cal D}h_{\mu\nu}\exp\left\{\int dx\left[-\frac1{12\eta^2}H_{\mu\nu
\lambda}^2+i\pi h_{\mu\nu}T_{\mu\nu}-\frac14 F_{\mu\nu}^2-\right.
\right.$$

$$-\left.\left.\varepsilon_
{\mu\nu\alpha\beta}F_{\mu\nu}\left(\frac{iNe}{4}h_{\alpha\beta}+
\frac{\pi}{Ne}T_{\alpha\beta}\right)\right]\right\}, \eqno (3) $$
where $H_{\mu\nu\lambda}\equiv\partial_\mu h_{\nu\lambda}+
\partial_\lambda h_{\mu\nu}+\partial_\nu h_{\lambda\mu}$ is a strength 
tensor of an antisymmetric field $h_{\mu\nu}$. From now on in this 
Section, we shall omit the preexponential factor, since our main result, 
the coefficient functions parametrizing the bilocal correlator of the 
dual field strength tensors (see Eqs. (11) and (12) below) will be 
obtained upon comparison of the two exponents, the result of the Gaussian 
integrations and the one of the cumulant expansion of Eq. (2).

One's first instinct is to fix some, for concreteness Feynman, gauge by 
adding to the Lagrangian standing in the exponent on the R.H.S. 
of Eq. (3) the gauge fixing term $\frac12\left(\partial_\mu 
A_\mu\right)^2$ and carry out the Gaussian integration 
over $A_\mu$, which then can be easily performed and 
yields 

$$\int {\cal D}A_\mu\exp\left\{-\int dx\left[\frac14 F_{\mu\nu}^2+
\varepsilon_{\mu\nu\alpha\beta}F_{\mu\nu}\left(\frac{iNe}{4}h_{\alpha
\beta}+\frac{\pi}{Ne}T_{\alpha\beta}\right)+\frac12 
\left(\partial_\mu A_\mu\right)^2\right]\right\}=$$

$$=\exp\left\{\frac{1}{\pi^2}\int dx dy\frac{1}{\left|x-y\right|^4}
\left[4\frac{(x-y)_\mu (x-y)_\nu}{(x-y)^2}-\delta_{\mu\nu}\right]\cdot
\right.$$

$$\left.\cdot\left[\left(\frac{2\pi}{Ne}\right)^2 T_{\lambda\mu}(x)
T_{\lambda\nu}(y)-
\left(\frac{Ne}{2}\right)^2 h_{\lambda\mu}(x) h_{\lambda\nu}(y)+2\pi i 
T_{\lambda\mu}(x)h_{\lambda\nu}(y)\right]\right\}. \eqno (4)$$ 
However we immediately see that Eqs. (3) and (4) provide us with 
such a representation of $\left<{\cal F}_N(S)\right>$, which is 
difficult to proceed with.

It is much more convenient to rewrite $\exp\left(-\frac14\int dx 
F_{\mu\nu}^2\right)$ as 

$$\int {\cal D} G_{\mu\nu}\exp\left\{\int dx
\left[-G_{\mu\nu}^2+\frac{i}{2}\varepsilon_{\mu\nu\alpha\beta}F_{\mu\nu}
G_{\alpha\beta}\right]\right\},$$
after which $A_\mu$- and $G_{\mu\nu}$-integrations yield

$$\left<{\cal F}_N(S)\right>=\int {\cal D}B_\mu 
{\cal D}h_{\mu\nu}\exp\left\{\int dx\left[-\frac{1}{12\eta^2}
H_{\mu\nu\lambda}^2+i\pi h_{\mu\nu}T_{\mu\nu}-\right.\right.$$

$$\left.\left.-\left(\frac{Ne}{2}
h_{\mu\nu}-\frac{2\pi i}{Ne}T_{\mu\nu}+\partial_\mu B_\nu-
\partial_\nu B_\mu\right)^2\right]\right\}, \eqno (5)$$
i.e. we have replaced the $A_\mu$-integration by the $B_\mu$ one. 
Next, performing the hypergauge transformation with the function $\frac{2}
{Ne}B_\mu$, which is equivalent to fixing the gauge by the condition 
$B_\mu=0$ (see the first Ref. in$^{8}$), we get from Eq. (5) 

$$\left<{\cal F}_N(S)\right>=\int {\cal D}
h_{\mu\nu}\exp \left\{\int dx\left[-\frac{1}{12\eta^2}H_{\mu\nu\lambda}^2
-\left(\frac{Ne}{2}\right)^2h_{\mu\nu}^2+\left(\frac{2\pi}{Ne}\right)^2
T_{\mu\nu}^2+3\pi i h_{\mu\nu}T_{\mu\nu}\right]\right\},\eqno (6)$$
which apart from the term 

$$\exp\left(\left(\frac{2\pi}{Ne}\right)^2
\int dx T_{\mu\nu}^2\right)=\exp\left(\left(\frac{2\pi}{Ne}\right)^2
\int\limits_S^{} 
d\sigma_{\mu\nu}(x)\int\limits_S^{} 
d\sigma_{\mu\nu}(x')\delta (x-x')\right) 
\eqno (7) $$
is the partition function of the massive Kalb-Ramond field 
interacting with the string (nonlinear massive axionic sigma model$^{12}$). 
Notice, that this is also a low-energy expression for the partition 
function of the 4D UCST, studied in Ref. 10, which is in fact the dual 
version of the 4D compact QED. Therefore one may conclude that at 
sufficiently large distances, when the term (7) (or, better to say, its 
regularized version, which will be presented immediately below) can 
be neglected, the Londons' limit of AHM coincides with the low-energy 
limit of the 4D compact QED. 

It is also worth mentioning, that the 
reduction of AHM in the Londons' limit to the nonlinear axionic 
sigma model enables one to apply methods developed for such models, 
for example, for the phenomenological description of fluctuations 
of the string world-sheet$^{13}$.  
  
Regularizing the term (7) and performing the   
Gaussian integration over $h_{\mu\nu}$ in Eq. (6) we arrive at the 
following representation for the weight factor (2)

$$\left<{\cal F}_N(S)\right>=\exp\left[\left(
\frac{m\eta}{2}\right)^2\int\limits_S^{} 
d\sigma_{\mu\nu}(x)\int\limits_S^{} d\sigma_{\mu\nu}
(x'){\rm e}^{-\frac{m^2}{4}(x-x')^2}-\right.$$

$$\left.-9\pi^2 \int\limits_S^{} 
d\sigma_{\lambda\nu} (x)\int\limits_S^{} d\sigma_{\mu\rho}(x')
D_{\lambda\nu, \mu\rho}(x-x')\right]. \eqno (8)$$
Here $m\equiv Ne\eta$ is the mass of the Kalb-Ramond field (equal 
to the mass of the gauge boson generated by the Higgs mechanism), which 
under the assumption that it is sufficiently large 
has been chosen as an UV cutoff for the term (7). The reason for such a 
choice of the UV cutoff is based on the large distance behaviour of the 
bilocal correlator, described by Eqs. (15) and (16) below, 
according to which the mass $m$ could be identified with the 
inverse correlation length of the vacuum in QCD, $T_g^{-1}$ 
(see also discussion in Ref. 13). The latter quantity is large 
in the confining regime w.r.t. $r^{-1}$, where $r$ is a typical 
length scale in the problem under study, i.e. a 
size of the Wilson loop in QCD, 
or a radius of the closed surface $S$ in our approach. This fact  
served as a motivation for the curvature expansion of the 
string effective action in Ref. 6, which is actually 
an expansion in powers of $\left(\frac{T_g}{r}\right)^2$. 

The propagator of the Kalb-Ramond field in Eq. (8) has the following form 
(the details of derivation of this propagator are presented in the 
Appendix to Ref. 13) 

$$D_{\lambda\nu, \mu\rho}(x)\equiv D_{\lambda\nu, \mu\rho}^{(1)}(x)+
D_{\lambda\nu, \mu\rho}^{(2)}(x),$$
where

$$D_{\lambda\nu, \mu\rho}^{(1)}(x)=\frac{Ne\eta^3}{8\pi^2}\frac
{K_1}{\left|x\right|}\Biggl(\delta_{\lambda\mu}\delta_{\nu\rho}-
\delta_{\mu\nu}\delta_{\lambda\rho}\Biggr), \eqno (9)$$

$$D_{\lambda\nu, \mu\rho}^{(2)}(x)=\frac{\eta}{4\pi^2Nex^2}\Biggl[\Biggl[
\frac{K_1}{\left|x\right|}+\frac m2\left(K_0+K_2\right)\Biggr]
\Biggl(\delta_{\lambda\mu}\delta_{\nu\rho}-\delta_{\mu\nu}\delta_{\lambda
\rho}\Biggr)+$$

$$+\frac{1}{2\left|x\right|}\Biggl[3\Biggl(\frac{m^2}{4}
+\frac{1}{x^2}\Biggr)K_1+\frac{3m}{2\left|x\right|}\left(K_0+
K_2\right)+\frac{m^2}{4}K_3\Biggr]\cdot$$

$$\cdot\Biggl(\delta_{\lambda\rho}x_\mu x_\nu+\delta_{\mu\nu}x_\lambda 
x_\rho-\delta_{\mu\lambda}x_\nu x_\rho-\delta_{\nu\rho}x_\mu x_\lambda
\Biggr)\Biggr], \eqno (10)$$
$K_i\equiv K_i(m\left|x\right|), i=0,1,2,3,$ stand for the Macdonald 
functions, and one can show that the term 

$$\int\limits_S^{} 
d\sigma_{\lambda\nu}(x)\int\limits_S^{} 
d\sigma_{\mu\rho}(x')D_{\lambda\nu, 
\mu\rho}^{(2)}(x-x')$$
could be rewritten as a boundary one, and therefore vanishes, since $S$ is 
closed. In particular, as it could be 
anticipated from the beginning, when $m\to 0$, the contribution 
of the function (9) vanishes, whereas 

$$\int\limits_S^{} 
d\sigma_{\lambda\nu}(x)\int\limits_S^{} 
d\sigma_{\mu\rho}(x')D_{\lambda\nu, 
\mu\rho}^{(2)}(x-x')\longrightarrow
\frac{1}{2\left(\pi Ne\right)^2}\oint
\limits_{\partial S}^{}dx_\mu\oint\limits_{\partial S}^{} dx'_\mu
\frac{1}{(x-x')^2}.$$

Comparing Eqs. (8)-(10) with the result of the cumulant 
expansion of the weight factor (2), we obtain the following 
values of the coefficient functions, which parametrize the bilocal 
correlator of the dual field strength tensors 

$$D\left(m^2x^2\right)=\frac{m^3}{16\pi^2}\left[9\frac{K_1}{\left|
x\right|}-m{\rm e}^{-\frac{m^2x^2}{4}}\right] \eqno (11)$$
and

$$D_1\left(m^2x^2\right)=\frac{9m}{8\pi^2x^2}\Biggl[\frac{K_1}{\left|
x\right|}+\frac{m}{2}\Biggl(K_0+K_2\Biggr)\Biggr], \eqno (12)$$
whose asymptotic behaviours at $\left|x\right|\ll\frac1m$ and $\left|
x\right|\gg\frac1m$ read 

$$D\longrightarrow\frac{9}{16\pi^2}\frac{m^2}{x^2}, \eqno (13)$$

$$D_1\longrightarrow\frac{9}{4\pi^2\left|x\right|^4} \eqno (14)$$
and

$$D\longrightarrow\frac{9m^4}{16\sqrt{2}\pi^{\frac32}}
\frac{{\rm e}^{-m\left|x\right|}}{\left(m\left|x\right|\right)^
{\frac32}}, \eqno (15)$$

$$D_1\longrightarrow\frac{9m^4}{8\sqrt{2}\pi^{\frac32}}
\frac{{\rm e}^{-m\left|x\right|}}{\left(m\left|x\right|\right)^
{\frac52}} \eqno (16)$$
respectively. 

We see that according to Eq. (8) all the terms of 
the cumulant expansion, higher than the quadratic one, vanish, so that the 
bilocal approximation to MVC is an exact 
result in the Londons' limit, or in another language, 
the ensemble of fields in this case is exactly 
Gaussian.  

The asymptotic behaviours (13)-(16) have already been discussed and 
compared with the lattice data$^{14}$ in 
Ref. 13. Namely, we see that Eqs. (15) and (16) 
reproduce correctly 
the long-range dynamics, i.e. the exponential fall-offs, and the result, 
that $D_1$ at large distances is smaller than $D$. 
However, as far as the short-distance 
dynamics is concerned, one can see that though Eq. (14) yields the 
correct $\frac{1}{\left|x\right|^4}$-behaviour of the function $D_1$, 
Eq. (13) agrees 
with the results obtained within the MVC in QCD in Ref. 15 only in 
the lowest order of perturbation 
theory, where the function $D$ is much smaller than the function $D_1$. 
This result 
is a consequence of the Abelian character of the model under study.

\vspace{6mm}
{\large \bf 3. Realistic 
Values of Charge and V.E.V. of the Higgs Field and Consistency of the 
Londons' Limit}
\vspace{3mm}

The derivative (or curvature, or $\frac{1}{m}$-) expansion of the action 

$$\left(\frac{m\eta}{2}\right)^2\int\limits_S^{} 
d\sigma_{\mu\nu}(x)\int\limits_S^{} d
\sigma_{\mu\nu}(x'){\rm e}^{-\frac{m^2}{4}(x-x')^2}-9\pi^2
\int\limits_S^{} 
d\sigma_{\lambda\nu}(x)\int\limits_S^{} 
d\sigma_{\mu\rho}(x')D_{\lambda\nu, 
\mu\rho}^{(1)}(x-x')$$ 
yields the following values of the string tension of the Nambu-Goto term 
and the inverse bare coupling constant of the rigidity term 

$$\sigma=\pi\eta^2\left(9K_0(Ne)-2\right)\simeq \pi\eta^2\left(9
\ln\frac{2}{\gamma Ne}-2\right)
\eqno (17)$$
and

$$\frac{1}{\alpha_0}=-\frac{5\pi}{8 \left(Ne\right)^2} \eqno (18)$$
respectively, where $\gamma\simeq 1.781$ stands for the Euler 
constant. These quantities are similar to the ones obtained for the 
low-energy expansion of the string effective action of the UCST in Ref. 10. 

We see that though the coupling constant (18) is always 
negative as 
it should be, the string tension (17) is positive only when the following 
inequality holds

$$Ne<\frac{2}{\gamma {\rm e}^{\frac29}}\simeq 0.899, \eqno (19)$$
i.e. only for sufficiently small values of the Higgs field charge. At this 
point it looks natural to address the important question on how exact 
is the Londons' limit of AHM in modelling the real confining regime of 
QCD? In order to answer this question, one should 
identify the mass $m$ with $T_g^{-1}$, 
and substitute the value of the latter in the 
$SU(3)$-gluodynamics$^{14}$, $T_g^{-1}\simeq 1{\mbox GeV}$, and the 
value of the string tension $\sigma\simeq 0.2 {\mbox GeV}^2$ (see, 
for example, Ref. 5) into Eq. (17). This yields 
the following equation 

$$Ne=1.12{\rm e}^{-0.22-0.01\left(Ne\right)^2},$$
whose solution according to MAPLE reads 

$$Ne\simeq 0.892, \eqno (20)$$
which luckly satisfies inequality (19). This means the consistency 
of the Londons' limit of AHM in modelling QCD in the confining regime. 
The corresponding square root of the Higgs v.e.v. reads

$$\eta\simeq 1.121 {\rm GeV}. \eqno (21)$$
Eqs. (20) and (21) provide us with the parameters of AHM in the Londons' 
limit, which according to the lattice data are the most adequate 
ones for the description of the confining phase of QCD. 

\vspace{6mm}
{\large \bf 4. Equation of Motion of the Weight Factor of the String 
Representation in the Loop Space}
\vspace{3mm}

In this short Section, we shall derive a loop equation for the weight 
factor (6) without the multiplier (7), which could be 
referred to the measure ${\cal D}x_\mu (\xi)$. We shall denote the weight 
factor (6) without this multiplier as $\left<{\bar 
{\cal F}}_N\left[x(\xi)\right]
\right>$.

Variation w.r.t. $h_{\nu\lambda}(x)$ under 
the functional integral yields

$$\Biggl<\Biggl(\frac{1}{2\eta^2}\Biggl(\Box h_{\nu\lambda}+
\partial_\mu\partial_\nu h_{\lambda\mu}+\partial_\mu
\partial_\lambda h_{\mu\nu}\Biggr)-\frac{\left(Ne\right)^2}{2}
h_{\nu\lambda}\Biggr)\Phi\left[h\left[x(\xi)\right]\right]+
3i\pi T_{\nu\lambda}\Phi\left[h\left[x(\xi)\right]\right]
\Biggr>_{h_{\mu\nu}}=0, \eqno (22)$$
where 

$$\Phi\left[h\left[x(\xi)\right]\right]\equiv\exp\Biggl(3i\pi
\int\limits_S^{} d\sigma_{\mu\nu}(x(\xi))h_{\mu\nu}\left[x(\xi)\right]
\Biggr),$$
and 
$$\left<...\right>_{h_{\mu\nu}}\equiv\int {\cal D}h_{\mu\nu}\exp\Biggl(-
\int dx\Biggl(\frac{1}{12\eta^2}H_{\mu\nu\lambda}^2+\left(\frac{Ne}
{2}\right)^2 h_{\mu\nu}^2\Biggr)\Biggr)\left(...\right).$$
Eq. (22) when being reformulated in the loop space reads 

$$\Biggl(\frac{1}{\eta^2}\Biggl(\partial^{x(\sigma){\,}2}
\frac{\delta}{\delta\sigma_{\nu\lambda}(x(\sigma))}+\partial_\mu^x
\partial_\nu^x\frac{\delta}{\delta\sigma_{\lambda\mu}(x)}+
\partial_\mu^x\partial_\lambda^x\frac{\delta}{\delta\sigma_{\mu\nu}(x)}
\Biggr)-$$

$$-\left(Ne\right)^2\frac{\delta}{\delta\sigma_{\nu\lambda}(x)}-
18\pi^2T_{\nu\lambda}\left[x(\sigma)\right]\Biggr)\left<{\bar {\cal F}}_N
\left[x(\sigma)
\right]\right>=0, \eqno (23)$$
where 
$x_\mu(\sigma), 0\le\sigma\le 1,$ is an element of the loop space, 
corresponding to an arbitrary closed contour $C$ lieing on $S$ [for 
example, $x_\mu(\sigma)=x_\mu\left(\xi_1,\xi_2={\rm const.}\right)$],  

$$\partial_\mu^{x(\sigma)}\equiv\int\limits_{\sigma-0}^{\sigma+0}
d\tau\frac{\delta}{\delta x_\mu(\tau)},$$
and 

$$\frac{\delta}{\delta\sigma_{\mu\nu}(x(\sigma))}=\int\limits_{-0}^
{+0}d\tau\tau\frac{\delta^2}{\delta x_\mu\left(\sigma+\frac12 \tau\right)
\delta x_\nu\left(\sigma-\frac12 \tau\right)}.$$
Eq. (23) coincides with the 
low-energy limit of the loop equation for the Wilson average in 4D 
compact QED, studied in Ref. 11, which is in 
the line with the discussion following after Eq. (7). The reader is 
referred to Ref. 11 for further investigations of Eq. (23).

\vspace{6mm}
{\large \bf 5. String Representation for the Partition Function 
of the Higgs Currents' Correlators}
\vspace{3mm}

Let us derive the bilocal correlator of the dual field strength tensors 
from the string representation for the 
partition function of an arbitrary number of the Higgs currents. In the 
Londons' limit, such a partition function reads

$${\cal Z}\left[J_\mu\right]=\frac{1}{{\cal Z}\left[0\right]}
\int {\cal D}A_\mu {\cal D}\theta\exp
\left\{\int dx\left[-\frac14 F_{\mu\nu}^2-\frac{\eta^2}{2}\left(
\partial_\mu\theta-NeA_\mu\right)^2+J_\mu j_\mu\right]\right\},$$
where $j_\mu\equiv -Ne\eta^2(\partial_\mu\theta-NeA_\mu)$ is the 
Higgs current in this limit. We use here the definition 
of ${\cal Z}\left[J_\mu\right]$ normalized to unity, since in what 
follows we shall get the function $D$ from the correlator 
$\left<j_\beta(x_1)j_\sigma(x_2)\right>$ itself, rather than from 
the comparison of two exponents, as it has been done in Section 2.  

Performing the same duality transformation as the one used in 
Section 2, we get the following string representation for ${\cal Z}
\left[J_\mu\right]$

$${\cal Z}\left[J_\mu\right]=\frac{{\rm e}^{\frac{m^2}{2}\int dx J^2(x)}}
{\int {\cal D}x_\mu (\xi){\rm e}^{-\pi^2\int\limits_S^{} 
d\sigma_{\lambda\nu}(x)
\int\limits_S^{} 
d\sigma_{\mu\rho}(x')D_{\lambda\nu, \mu\rho}(x-x')}}
\int {\cal D}x_\mu (\xi){\rm e}^{-\pi^2\int\limits_S^{} 
d\sigma_{\lambda\nu}(x)
\int\limits_S^{} 
d\sigma_{\mu\rho}(x')D_{\lambda\nu, \mu\rho}(x-x')}\cdot$$

$$\cdot\exp\Biggl\{Ne\varepsilon_{\lambda\nu\alpha\beta}\int dx dy\Biggl[
-\frac{Ne}{4}\varepsilon_{\mu\rho\gamma\delta}\Biggl(\frac{\partial^2}
{\partial x_\alpha\partial y_\gamma}D_{\lambda\nu, \mu\rho}(x-y)
\Biggr)J_\beta (x) J_\delta (y)+$$

$$+\pi T_{\mu\rho}(y)\Biggl(
\frac{\partial}{\partial x_\alpha} D_{\lambda\nu, \mu\rho}(x-y)\Biggr)
J_\beta (x)\Biggr]\Biggr\}. \eqno (24)$$
Upon the two-fold variation of Eq. (24) w.r.t. $J_\mu$, we arrive 
at the following expression for the correlator of two Higgs currents  

$$\left<j_\beta(x_1) j_\sigma(x_2)\right>=\frac{\left(Ne\right)^2}
{\int {\cal D}x_\mu(\xi)
{\rm e}^{-\pi^2\int\limits_S^{} 
d\sigma_{\lambda\nu}(x)\int\limits_S^{} d\sigma_{\mu\rho}
(x')D_{\lambda\nu, \mu\rho}(x-x')}}\cdot$$

$$\cdot\int {\cal D}x_\mu(\xi)
{\rm e}^{-\pi^2\int\limits_S^{} 
d\sigma_{\lambda\nu}(x)\int\limits_S^{} d\sigma_{\mu\rho}
(x')D_{\lambda\nu, \mu\rho}(x-x')}\Biggl\{\eta^2\delta_{\beta\sigma}
\delta(x_1-x_2)+$$

$$+\varepsilon_{\lambda\nu\alpha\beta}\varepsilon_
{\mu\rho\gamma\sigma}\Biggl[-\frac12\frac{\partial^2}{\partial x_{1_
\alpha} \partial x_{2_\gamma}}D_{\lambda\nu, \mu\rho}(x_1-x_2)+$$

$$+\pi^2
\int\limits_S^{} d\sigma_{\delta\zeta}(y)\int\limits_S^{} 
d\sigma_{\chi\varphi}(z)
\Biggl(\frac{\partial}{\partial x_{1_\alpha}}D_{\lambda\nu, 
\delta\zeta}(x_1-y)\Biggr)\Biggl(\frac{\partial}
{\partial x_{2_\gamma}} D_{\mu\rho, \chi\varphi}(x_2-z)\Biggr)\Biggr]
\Biggr\}. \eqno (25)$$

The first term in the figured brackets on the R.H.S. of Eq. (25) obviously 
vanishes for 
$x_1\ne x_2$. It is also easy to show that the contribution of the 
term (10) to the R.H.S. of Eq. (25) vanishes too. As far as the 
part of the axionic propagator given by Eq. (9), 
is concerned, one can see that it   
has the order of magnitude of $\left(Ne\right)^2\eta^4$.  
Let us now assume that $Ne\eta^2\left|S\right|\ll 1$, 
where $\left|S\right|$ stands for the area of the surface $S$, so that 
the second term in square brackets 
on the R.H.S. of Eq. (25) is much smaller than the 
first term, and could be disregarded. This approximation corresponds 
to the case of a very small $\eta$, i.e. to a very narrow potential. 

Taking all this into account and making use of the 
equation$^{5}$,

$$\left<j_\beta(x_1)j_\sigma(x_2)\right>=\Biggl(\frac{\partial^2}
{\partial x_{1_\mu}\partial x_{2_\mu}}\delta_{\beta\sigma}-
\frac{\partial^2}{\partial x_{1_\beta}\partial x_{2_\sigma}}\Biggr)
D\left(m^2\left(x_1-x_2\right)^2\right),$$
which could be obtained from the classical equation of motion, 
we get the following (semiclassical) value of the function $D$

$$D\left(m^2x^2\right)=\frac{m^3}{4\pi^2}\frac{K_1
\left(m\left|x\right|\right)}
{\left|x\right|},~ x\ne 0. \eqno (26)$$
One can see that Eq. (26) differs in a factor $\frac94$ from 
the leading-order 
short-distance dynamics of Eq. (11), given by Eq. (13) (or from Eq. (11) 
itself at $x\ne 0$, where the last term on the R.H.S. 
without regularization 
vanishes). Besides the 
neglection of the second term in square brackets 
on the R.H.S. of Eq. (25), the difference 
between $\frac94$ and $1$ describes the error of the semiclassical 
approximation.  
Notice also, that only 
the function $D$ could be obtained from the correlator (25) in the 
semiclassical approximation due to 
the independence of the latter of the function $D_1$ 
in this approximation.

\vspace{6mm}
{\large \bf 6. Conclusion}
\vspace{3mm}

Let us now stress some physical consequences of the obtained results 
and discuss open questions. The first outcome of the 
Letter is the result of Section 2 that at large distances AHM in 
the Londons' 
limit coincides with the low-energy limit of the 4D compact QED. 
Secondly, we have demonstrated that the bilocal approximation to the 
MVC is exact in the Londons' limit, i.e. all the higher cumulants 
vanish. Thirdly, we have proved that 
the Londons' limit is consistent for the description of 
the confining phase 
of QCD according to the lattice data, and found the most adequate to 
QCD values 
of charge and the square root of the v.e.v. of the Higgs field, which are 
given by Eqs. (20) and (21) respectively. The most important result of the 
Letter is the two scalar functions, $D$ and $D_1$, given by Eqs. (11) and 
(12) respectively, which parametrize 
the correlator of the two dual field strength tensors and are known to 
be the key quantities in the MVC. 

Notice, that already in Ref. 13 it has 
been conjectured that the functions $D$ and $D_1$ must follow from 
the propagator 
of the massive Kalb-Ramond field. In Ref. 13, this conjecture has been 
supported by comparison of the asymptotic behaviours of these functions 
at small and large distances obtained from such a propagator 
with the ones observed in the lattice measurements$^{14}$. In this 
Letter, we have proved this conjecture and in addition obtained  
integration over metrics resulting from the integration over the 
singular part of the Higgs wave function. However, it is unclear what 
collective variable in QCD may serve for getting such an integration 
over metrics. As it has been already mentioned in the Introduction, the 
most physically appropriate candidate for this variable is the 
strong background fields, which provide confinement in the QCD vacuum. 
The solution of the problem of whether this is really the case or not 
will be the topic of a separate publication.

It is also worth mentioning, that, 
as it has been discussed in Section 2, the short-distance 
asymptotics (13) of the function $D$ agrees with the MVC only in the 
lowest order of perturbation theory. In order to get better 
correspondence with the MVC, one should investigate AHM in the vicinity 
of the Londons' limit, i.e. account for the corrections which come 
from the finiteness of the coupling constant $\lambda$ in Eq. (1), 
which presumably takes place for the ``physical'' AHM, i.e. the one, 
which describes QCD with the best accuracy$^{16}$. In this way, the bilocal 
approximation will not be exact any more, and higher cumulants will 
start contributing into the Nambu-Goto term string tension and 
the rigidity term coupling constant. The effects arising due to these 
cumulants will be studied in the next publications.  

The final (technical) problem, which must be solved, is accounting for   
the Jacobian, which arises when one passes from the integration over 
the singular part of the Higgs wave function to the integration over 
metrics and has been omitted for simplicity in this Letter. This 
Jacobian has been calculated in Ref. 1 and, as it has been mentioned 
in the Introduction, yields the Polchinski-Strominger term$^{3}$ with 
the coupling constant ensuring exact cancellation of the conformal 
anomaly in D=4.

\vspace{6mm}
{\large \bf 7. Acknowledgments}
\vspace{3mm}

The author is deeply grateful to Profs. Yu.M.Makeenko, M.I.Polikarpov 
and Yu.A.Simonov, and also to E.T.Akhmedov for the 
useful discussions and encouragement. He is also indebted to M.N.Chernodub 
for critical reading the manuscript and a lot of useful remarks.  
He would also like to thank the theory group of the 
Quantum Field Theory Department of the Institute of Physics of the 
Humboldt University of Berlin and especially Profs. D.Ebert, 
D.L\"ust, and M.M\"uller-Preu{\ss}ker for kind hospitality.

\vspace{6mm}
{\large \bf References}

\vspace{3mm}
\noindent
1.~M.I.Polikarpov, U.-J.Wiese, and M.A.Zubkov, {\it Phys.Lett.} {\bf B309}, 
133 (1993); 
E.T.Akhmedov, M.N.Chernodub, M.I.Polikarpov, and M.A.Zubkov, 
{\it Phys.Rev.} 
{\bf D53}, 2087 (1996); E.T.Akhmedov, {\it JETP Lett.} 
{\bf 64}, 82 (1996).\\ 
2.~K.Lee, {\it Phys.Rev.} {\bf D48}, 2493 (1993).\\
3.~J.Polchinski and A.Strominger, {\it Phys.Rev.Lett.} {\bf 67}, 
1681 (1991).\\
4.~H.G.Dosch, {\it Phys.Lett.} {\bf 190}, 177 (1987); Yu.A.Simonov, 
{\it Nucl.Phys.} 
{\bf B307}, 512 (1988); H.G.Dosch and Yu.A.Simonov, {\it Phys.Lett.} 
{\bf B205}, 
339 (1988), {\it Z.Phys.} {\bf C45}, 147 (1989); Yu.A.Simonov, 
{\it Nucl.Phys.}  
{\bf B324}, 67 (1989), {\it Phys.Lett.} {\bf B226}, 151 (1989), 
{\it Phys.Lett.}  
{\bf B228}, 413 (1989), {\it Yad.Fiz.} {\bf 58}, 113 (1995), preprint 
ITEP-PH-97-4 
({\it hep-ph}/9704301); for a review see   
Yu.A.Simonov, {\it Yad.Fiz.} {\bf 54}, 192 (1991).\\
5.~Yu.A.Simonov {\it Phys.Usp.}  
{\bf 39}, 313 (1996).\\
6.~D.V.Antonov, D.Ebert, and Yu.A.Simonov, {\it Mod.Phys.Lett.} {\bf A11}, 
1905 (1996).\\
7.~H.Kleinert, {\it Phys.Lett.} {\bf B211}, 151 (1988); K.I.Maeda and 
N.Turok, {\it Phys.Lett.} {\bf B202}, 376 (1988); S.M.Barr and D.Hochberg, 
{\it Phys.Rev.} {\bf D39}, 2308 (1989); H.Kleinert, {\it Phys.Lett.} 
{\bf B246}, 127 (1990); 
H.Kleinert and 
A.M.Chervyakov, {\it hep-th}/9601030 (in press in {\it Phys.Lett.} 
{\bf B}); M.Anderson, F.Bonjour, R.Gregory, and J.Stewart, 
{\it Phys.Rev.} {\bf D56}, 8014 (1997).\\
8.~P.Orland, {\it Nucl.Phys.} {\bf B428}, 221 (1994); 
M.Sato and S.Yahikozawa, {\it Nucl.Phys.} {\bf B436}, 100 (1995).\\
9.~A.M.Polyakov, {\it Nucl.Phys.} {\bf B486}, 23 (1997).\\
10.M.C.Diamantini, F.Quevedo, and C.A.Trugenberger, {\it Phys.Lett.}  
{\bf B396}, 115 (1997).\\
11.D.V.Antonov, {\it Mod.Phys.Lett.} {\bf A12}, 1419 (1997).\\ 
12.M.B.Green, J.H.Schwarz, and E.Witten, {\it Superstring Theory}, 
Vols. I, II  
(Cambridge University Press, England, 1987).\\
13.D.V.Antonov, {\it hep-th}/9707245.\\
14.A. Di Giacomo and H.Panagopoulos, {\it Phys.Lett.} {\bf B285}, 
133 (1992); 
A. Di Giacomo, E.Meggiolaro, and H.Panagopoulos, {\it hep-lat}/9603017 
(preprints IFUP-TH 12/96 and UCY-PHY-96/5) (in press in {\it Nucl.Phys.} 
{\bf B}).\\
15. M.Eidem\"uller and M.Jamin, {\it hep-ph}/9709419; Yu.A.Simonov and 
V.I.Shevchenko, in preparation.\\
16. S.Kato, M.N.Chernodub, S.Kitahara, N.Nakamura, M.I.Polikarpov, and 
T.Suzuki, preprint KANAZAWA-97-15 ({\it hep-lat}/9709092); 
M.N.Chernodub and M.I.Polikarpov, preprint ITEP-TH-55-97 
({\it hep-th}/9710205).

\end{document}